\documentclass[aps,twocolumn,pra,notitlepage,]{revtex4-2}
\usepackage{amsmath,amssymb}
\usepackage{dsfont}
\usepackage{graphicx}
\usepackage{physics}
\usepackage{mathtools}
\usepackage{cancel}
\usepackage{bbold}
\usepackage{bm}
\usepackage{color} 
\usepackage[dvipsnames]{xcolor}
\usepackage[T1]{fontenc}
\usepackage[colorlinks,urlcolor=blue,linkcolor=blue,citecolor=blue]{hyperref}
\usepackage{tcolorbox}

\usepackage{makecell} 
\usepackage{multirow} 

\usepackage{soul}

\usepackage{ulem} 

\begin{abstract}
    Properties of quantum dot based spin qubits have significant inter-device variability due to unavoidable presence of various types of disorder in semiconductor nanostructures. A significant source of this variability is charge disorder at the semiconductor-oxide interface, which causes unpredictable, yet, as we show here, correlated fluctuations in such essential properties of quantum dots like their mutual tunnel couplings, and electronic confinement energies. This study presents a systematic approach to characterize and mitigate the effects of such disorder. We utilize finite element modeling of a Si/SiGe double quantum dot to generate a large statistical ensemble of devices, simulating the impact of trapped interface charges. This work results in a predictive statistical model capable of generating realistic artificial data for training machine learning algorithms. By applying Principal Component Analysis to this dataset, we identify the dominant modes through which disorder affects the multi-dimensional parameter space of the device. Our findings show that the parameter variations are not arbitrary, but are concentrated along a few principal axes -- i.e.~there are significant correlations between many properties of the devices. 
    We finally compare that against control modes generated by sweeping the gate voltages, revealing limitations of the plunger-only control. This work provides a framework for enhancing the controllability and operational yield of spin qubit devices, by systematically addressing the nature of electrostatic disorder that leads to statistical  correlations in properties of double quantum dots.
\end{abstract}

\graphicspath{{figures_var/}}

\begin{document}
\newcounter{theo}

\author{Saeed Samadi}
\affiliation{Institute of Physics, Polish Academy of Sciences, Poland}

\author{Łukasz Cywiński}
\affiliation{Institute of Physics, Polish Academy of Sciences, Poland}
\author{Jan A. Krzywda}\email{j.a.krzywda@liacs.leidenuniv.nl}
\affiliation{$\langle aQa^L
\rangle$ Applied Quantum Algorithms, Lorentz Institute and Leiden Institute of Advanced Computer Science,
Leiden University, The Netherlands}

\title{Statistical Structure of Charge Disorder in Si/SiGe Quantum Dots} 

\maketitle

\section{Introduction}

Spin qubit devices based on quantum dots in Si/SiGe nanostructures already offer high-fidelity single-qubit and two-qubit gates \cite{Kawakami_PNAS16,Yoneda_NN18,Zajac_Science18,Watson_Nature18,Xue_Nature22,Noiri_Nature22,Mills_SA22,Wu_arXiv25}, and readout \cite{Connors_PRAPL20,Noiri_NL20}, while maintaining relatively long coherence times \cite{Yoneda_NN18,Struck_NPJQI20}, and shuttling-based connectivity \cite{Mills_NC19,Seidler_NPJQI22,Zwerver_PRXQ23,Xue_NC24,Struck_NC24,DeSmet_NN25}. This makes them a viable platform for both digital \cite{Philips_Nature22,Fuentes_arXiv25}  and analog quantum computing \cite{Hensgens2017_AnalogSim, Dehollain2020_Nagaoka}. 

The key long-term advantage of quantum computing architecture based on silicon quantum dots is the possibility of leveraging industrial manufacturing technologies \cite{Zwerver_NE22,Neyens_Nature24,George_NL25,Huckemann_IEEE25} to create devices containing millions of qubits on a chip that can easily fit into a standard dilution refrigerator. However, variability of properties of such qubits realized in a solid-state technology remains a challenge for quantum computing architecture \cite{Vandersypen_NPJQI17}, as much more attention has to be given to tuning of single-qubit properties, and parameters of two-qubit gates than in case of qubits based on identical atoms or ions \cite{Bruzewicz_APR19, Henriet_Quantum20}. The presence of large variability of parameters of quantum dot devices created according to the same blueprint on the same heterostructure has been an experimental challenge since the beginning of research into quantum dot based spin qubits. However, only recently developments leveraging industrial technology have allowed for making measurements on large numbers of quantum dots defined on the same wafer \cite{Neyens_Nature24}, leading to a qualitative change in the amount of data on variability of properties of qubits created in a given structure. Such developments provide motivation for calculations of statistical distributions of various parameters of qubits for given microscopic models of disorder in nanostructures.

Variability of properties of qubits in Si/SiGe quantum dots can be caused by electrostatic disorder due to the presence of charged impurities and defects at random positions in the nanostructure \cite{Klos_PRB18,Martinez_PRAPL24}, atomistic disorder and interface roughness at Si/SiGe interface \cite{Martinez_PRAPL22,Wuetz_NC22,Losert_PRB23,Pena_NPJQI24} and inhomogeneous strain \cite{Wiciak_PRAPL23}. These effects were also discussed theoretically for SiMOS devices \cite{Martinez_PRAPL22,Cifuentes_NC24}, but the degree to which their conclusions from these works apply to Si/SiGe (for which the distance between the qubits and charge traps in an insulator, and the nature of interface roughness are distinct from SiMOS) is unclear. 

Due to large variability of the basic parameters of devices hosting multiple quantum dots, such as on-site energies and interdot tunnel couplings, the voltages on the gates defining the multi-dot system have to be tuned in order to even start considering the system as a multi-qubit register. For registers consisting of $N>2$ qubits this tuning quickly becomes unmanageable with increasing $N$ if done manually, and it has to be automated \cite{Zwolak_PRAPL20,Zwolak_RMP23}. Machine learning algorithms used for these purposes have to be trained on simulated data on response of disordered multi-dot devices to control voltages \cite{craig2024bridging}. Identifying the realistic microscopic model of disorder that captures the major features of qubit variability would allow then for generation of more realistic training data.

In this paper, we focus on electrostatic disorder due to density $\rho$ of charges trapped at SiGe/SiO$_{2}$ interface, ranging from $5 \times 10^{9}$ to $5 \times 10^{10}\,\text{cm}^{-2}$ \cite{Klos_PRB18,Martinez_PRAPL24,Kepa_charge_APL23}, and consider the variability of parameters of a double quantum dot (DQD) caused by it.
 Using finite element modelling of a realistic Si/SiGe double quantum dot with trapped interface charges, we generate a statistical ensemble of devices.
The DQD is controlled by two plunger gates, and also possibly by a barrier gate \cite{Martins_PRL16,Reed_PRL16}. At given $\rho$, for every realization of disorder, we tune the voltages on these gates to identify the range of voltage detunings at which a single electron is transferred between the two dots and, in this way, obtain a functional DQD device. The detuning needed to bring the DQD into a symmetric shape at which the charge transition occurs is the first quantity on the statistics of which we focus. We then reconstruct the statistical distribution of many other quantities: interdot tunnel coupling $t_c$, interdot distance $d$, interdot barrier height $h_B$, orbital energies and confinement lengths in both dots, and disorder-induced electric fields in the $z$ direction acting on both QDs. Next we move beyond quantifying the magnitude of parameter variations, and uncover the characteristic 'modes' of disorder. We demonstrate that disorder modifies parameters of DQDs along specific directions in the parameter space, and certain parameters exhibit significant correlations. An interesting example is a strong correlation between interdot distance $d$ and interdot barrier height $h_B$, the presence of which allows us to identify a robust quantitative relation between $d$ and $t_c$.

To uncover and quantitatively analyze the characteristic modes of disorder, and to further assess their controllability, we employ Principal Component Analysis (PCA) \cite{Carleo_RMP19}, which is a dimensionality reduction method based on a spectral analysis of the covariance matrix.
Within the spin qubit literature, the application of PCA has been largely limited to technical pre-processing steps, such as improving the signal-to-noise ratio of measurement data \cite{Wuetz_NC23}. In contrast, here we use it to identify the principal directions in the parameter space along which fluctuations occur (the eigenvectors) and to quantify their corresponding variance (the eigenvalues). When the variance is concentrated in just a few dominant components, one can effectively truncate the parameter space and perform the analysis within a low-dimensional manifold \cite{bhakuni2024diagnosing}. In our case, this manifold captures the essential physics of the electrostatic disorder, and allows us to build a resource-efficient predictive statistical model of a disordered DQD using data from microscopic simulations.
We have applied the PCA method to our simulated disorder data and found three statistically relevant modes that correspond to characteristic features of spatial realizations of charge disorder.
The mode responsible for the most variability, for instance, is caused by fluctuations of number of charged defects located {\it between} the two dots. By comparing the manifold spanned by these disorder modes to the one spanned by the control modes, we found that the effects of this dominant mode cannot be reversed without the use of the barrier gate. This framework thus provides a clear, quantitative demonstration of the limitations of plunger-only control scheme \cite{Ivlev2025}, in mitigating realistic device-to-device variability.
Our work establishes thus PCA as a primary framework for physical interpretation, using it to deconstruct the structure of device-to-device variability for the microscopic model of disorder discussed here, and to rigorously quantify system controllability.

The paper is structured as follows. Section~\ref{sec:modeling} introduces our device model. We then quantify the device-to-device variability in Sec.~\ref{sec:variability}, first by its impact on functional device yield (defined by thresholds for amount of correction to detuning needed to obtain a symmetric DQD, and values of orbital energies and tunnel couplings of the resulting device), 
and then by analyzing the underlying fluctuations in tunnel coupling. To deconstruct the root causes of this variability, in Sec.~\ref{sec:PCA} we use the principal component analysis to identify the dominant disorder modes, and build a predictive statistical model of disorder. This framework is then applied in Sec.~\ref{sec:control} to quantify the effectiveness of gate control in mitigating these effects, comparing standard two-gate and three-gate schemes. We conclude with a discussion in Sec.~\ref{sec:discussion}. Appendix~\ref{app:pca} details the supplementary results of PCA and DQD parameter fluctuations, while Appendix~\ref{app:notuning} provides the results for controllability, all relating to the alternative operational gate voltage setting.

\section{Device Model and Simulation \label{sec:modeling}}
\begin{figure*}[htp!]
    \centering
    \includegraphics[width=0.99\textwidth]{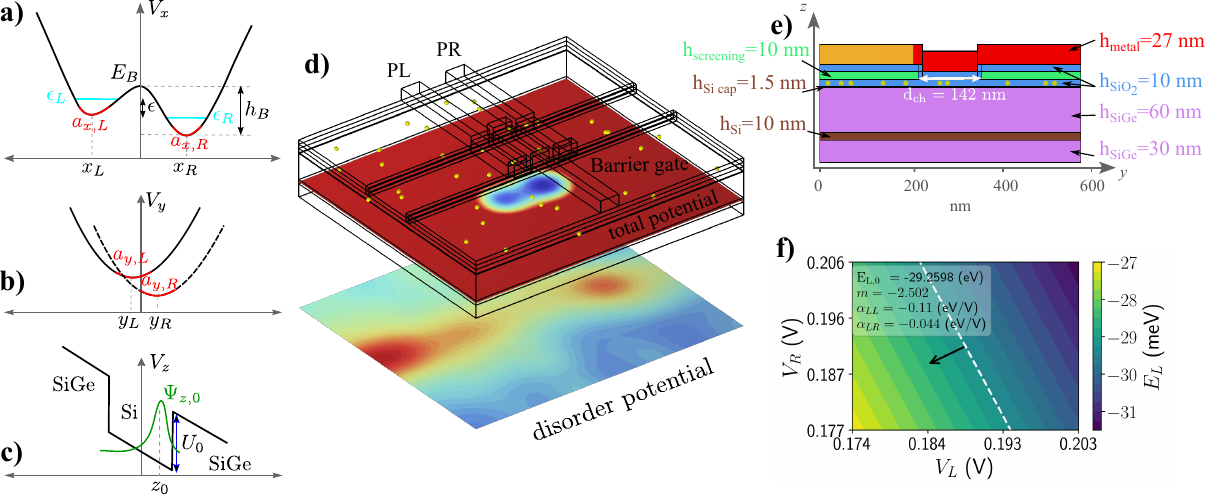}
    \caption{\textbf{Device Model and Experimental Regime.} (a)-(c) schematic diagrams of double-well potential along three axes used to present relevant DQD parameters (d) Schematic of the simulated Si/SiGe DQD device structure showing the gate layout and trapped charges. Example distribution of the disorder potential generated solely by defects, shown with all gates turned off. e) side view of DQD device  (f) Extracted plunger gate lever arm $\alpha$, confirming agreement with typical experimental values.  }
    \label{fig:device}
\end{figure*}

We model the double quantum dot using the finite element method, as implemented in COMSOL Multiphysics \cite{comsol}. This allows us to determine the electrostatic confinement potential used in the single-electron Schrödinger equation. The model, illustrated in Fig.~\ref{fig:device}, includes metallic top gates and spatially random distribution of fixed charges at the SiO$_2$/Si cap interface having density $\rho$. Our simulated heterostructure is grown along the $\hat{z}$ axis corresponding to [001] crystallographic direction, and consists of a Si quantum well between two SiGe barriers, the Si cap, two insulating regions (SiO$_2$), and a screening region. 
Metallic gates are employed to trap and confine electrons within the Si layer, and to induce an external electric field along the $\hat{z}$ direction, thereby enabling tunability of the electronic states. 

For concreteness, we analyze a quantum well of thickness $h_{\text{Si}} = 10 \,\text{nm}$ \cite{Struck_NPJQI20,liu2023growth}. However, it should be noted that with the applied electric field in $z$ direction $F_z \! =\! 5$ MV/m the electron wavefunction is localized within $\approx \!5$ nm from the top Si/SiGe interface, and the exact value of $h_{\text{Si}}$ that is larger than this localization length is irrelevant when considering the effects of electrostatic disorder on electronic states. Hence our calculations apply also to the case of thinner quantum wells, such as those with $h_{\text{Si}} \!=\! 5$ nm chosen to below the Matthews-Blakeslee
critical thickness for strain relaxation \cite{liu2022role,Wuetz_NC23}.

The interface between the upper SiGe barrier and the insulating region is at $h_{\text{top}} = 60 \,\text{nm}$, distance from the Si/SIGe interface, and it is followed by an additional 1.5 nm of Si cap layer. We use a relative permittivity of \( \epsilon_r = 3.9 \) for the SiO$_2$ oxide  and $\epsilon_{\mathrm{Si}}\!= \!12$, $\epsilon_{\mathrm{SiGe}}\! = \! 13.2$. Following prior Si/SiGe quantum-dot modeling, we take a conduction-band offset of \( U_0 = 150~\mathrm{meV} \) for Si well with Si\(_{0.7}\)Ge\(_{0.3}\) barriers
 \cite{Lima_MQT23}. In the insulating region, we impose $U \to \infty$ (or a hard-wall condition in the COMSOL solver) to ensure electron confinement. This framework captures the essential physics of electron confinement in  Si/SiGe quantum dots, including the anisotropy of the effective masses, gate-induced fields, lateral ellipticity, and heterostructure band offsets.

The device structure shown in Fig.~\ref{fig:device}~(d) has overall dimensions of \(660 \times 582~\text{nm}^2\). 
The channel is defined as a gap in the screening layer, with a width of \(d_{\text{channel}} = 142~\text{nm}\). 
The metallic gates have a width of \(w_{\text{metal}} = 45~\text{nm}\) and a height of \(h_{\text{metal}} = 27~\text{nm}\). 
The remaining dimensions are provided in Fig.~\ref{fig:device}(e).
We approximate the effectively two-dimensional potential that leads to lateral confinement by evaluating the numerically obtained three-dimensional potential $V(x,y,z)$ at $z$ corresponding to expectation value of this coordinate for an electron confined in the quantum well.
The numerically obtained lateral confinement is then fitted near the two potential minima by harmonic potentials characterized by $a_{x(y)} \! =\!  m_t \omega_{x(y)}^2$ curvatures, where $\hbar\omega_{x(y)}$ are the energies of quantum harmonic oscillator,  $m_t = 0.19 m_e$ is the relevant effective mass, and the resulting characteristic in-plane confinement lengths are $ L_{x(y)} \!= \!( \hbar^2/m_t a_{x(y)})^{1/4}$. 

In the absence of charged defects, we tune the device to achieve a tunneling gap of \( 2 t_{c,0} \approx 24\,\mu\text{eV} \) and an interdot distance of \( d_0 = 95\,\text{nm} \), corresponding to a slightly elliptical quantum dot with lateral dimensions \( L_{x,0} = 20\,\text{nm} \) and \( L_{y.0} = 18\,\text{nm} \), giving a ratio of $r_{L/R}= L_{x,0}/L_{y,0} \approx 1.08$.  These parameters are both realistic and experimentally relevant \cite{mcjunkin2022sige, Raith2011lateral}. 
The energy contribution from the applied vertical electric field is described by the term \( -e F_z z \), where the field strength in defect-free DQD is \( F_{z,0} \! =\!  5.34\,\text{MV/m} \) 
.  To achieve this, we set plunger gates 
\( V_{L/R} = 0.35\,\mathrm{V} \) to obtain an orbital excitation energy of 
\( E_{\mathrm{orb}} \approx 1.5\,\mathrm{meV} \) in each dot, with the barrier gate set to 
\( V_B = -0.1\,\mathrm{V} \).  

These values have been chosen to guarantee that in presence of $\rho \! =\! 5\times 10^{10}$ cm$^{-2}$ charges at the semiconductor/oxide interface the electrostatic potential in the Si quantum well still possesses two local minima, i.e.~the DQD can be still found in the presence of disorder.

In such a gate-defined quantum dot, the energy at the left potential minimum, $E_L$, can be written to a very good approximation as a linear function of plunger voltages,
\begin{equation}
E_L(V_L, V_R) = \alpha_{LL} V_L + \alpha_{LR} V_R + c_L \,\, ,
\end{equation}
and we determine the lever arms $\alpha_{LL}, \alpha_{LR}$ by ordinary least squares
fit to the calculated $(V_L,V_R,E_L)$ data. Lever arms quantify how efficiently each physical gate shifts the dot energy—information used to construct virtual gates \cite{volk2019loading,Hsiao2020Efficient,Ziegler2023Automated} that independently control detuning and occupancy while minimizing cross-talk, thereby enabling precise qubit calibration. The constant-energy contours (charge transition lines) in the $(V_L,V_R)$ plane satisfy $E_L = \text{const}$ and thus have slope \(
m = -\alpha_{LL}/\alpha_{LR}
\).
The energy gradient line in gate space is orthogonal to the contours and indicating the gate combination that most strongly shifts $E_L$. Extracted lever-arm parameters, defining constant-energy contours and energy gradients in $(V_L,V_R)$ space for the defect-free device, are shown in Fig.~\ref{fig:device}(f). Our analysis shows that the charge disorder induced variation in the lever arm remains below 10\% across the considered density range. As a result, we have chosen to exclude it from further investigations.

To characterize the DQD in presence of a given spatial realization of disorder (positions of charges at the interface), we first set the gate voltages to the previously optimized values, 
and then tune the plunger gates to locate the tunnel anticrossing. The observation of this anticrossing confirms the formation of a DQD. Once the anticrossing is identified, the in-plane potential is extracted and fitted to its two minima, as illustrated in Fig.~\ref{fig:device}~(a).
We begin by fitting the most relevant parameters of the double-well potential—\(E_D\) (the minimum of the parabola for dot \(D\)), the dot position \((x_D, y_D)\), and the curvatures \(a_{{x(y)},D}\). The single-dot lateral confinement energy is \(
\epsilon_D = E_D + \hbar(\omega_{x,D}+\omega_{y,D})/{2}.
\)
The value of the electric field \( F_{z,D} \), with or without impurities, is obtained at each dot position \( D \).

In presence of interface charges the two minima of the potential obtained for initial plunger voltages are detuned. We obtain this detuning by finding the voltage difference that makes the DQD symmetric, and using the lever arm to convert it into the disorder-induced energy detuning:
\begin{equation}
\epsilon = \alpha \,\frac{\Delta V_{LR}}{2}, \qquad
\alpha = \alpha_{LL} + \alpha_{RR} - \alpha_{LR} - \alpha_{RL}.
\end{equation}
For the disorder-free case where the value of $|\alpha|$ is 0.12 eV/V. In presence of $\rho \! \sim \! 10^{10}$ cm$^{-2}$ defect charges the variation of $\alpha$ is $<\! 10 \%$. 

Among the many parameters involved, we focus on a subset that mainly governs DQD qubit operations. This parameter vector, which forms the basis of our statistical analysis, includes the tunnel coupling $t_c$ (tunneling gap \(2t_c\)), the interdot distance \(d\), the interdot barrier height \(h_B\), the disorder-induced energy detuning \(\epsilon\), the average confinement length \(L_x\) along the axis connecting the two dots and its left–right difference \(\Delta L_x\), and the average vertical electric field \(F_z\) together with its left–right difference \(\Delta F_z\), and the confinement shape ratio $r_D = L_x / L_y$ at dot $D$.

\section{\label{sec:variability} Quantifying Device-to-Device Variability}
In this section, we quantify the variability of the double quantum dot by analyzing an ensemble of devices subject to independent realizations of charge disorder. This model aims to reconstruct device-to-device, or thermal-cycle variability observed in the experiments \cite{craig2024bridging}. For each instance of disorder, we tune the dots to 
zero energy detuning before extracting their final parameters. This ensures we are analyzing variability around a consistent operational point.

\subsection{Impact on Reproducibility}

\begin{figure}[htb!]
    \centering
    \includegraphics[width=0.9\columnwidth]{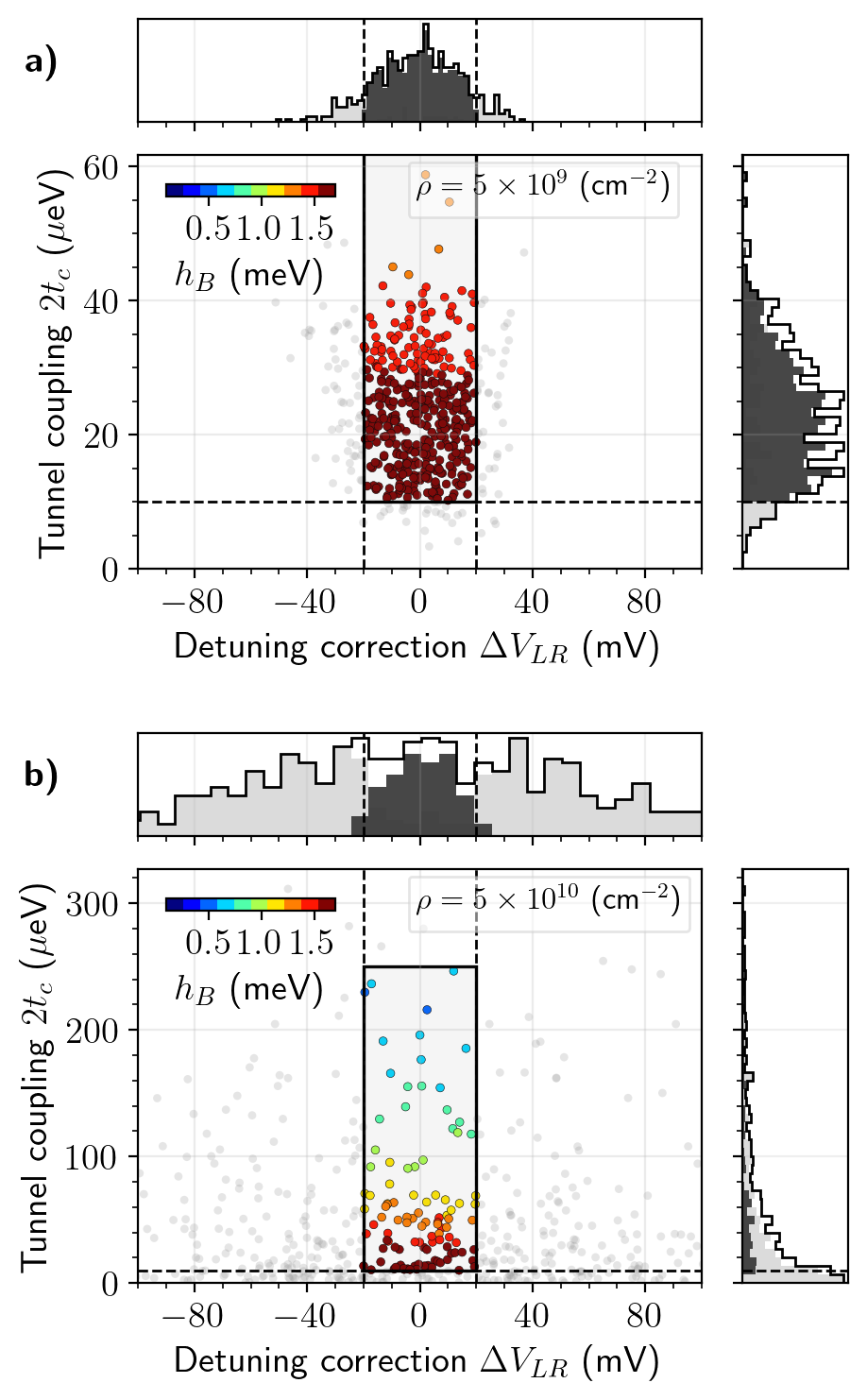}
    \caption{Estimated yields of 76\% and 20\% are obtained for (a) $\rho = 5 \times 10^{9}$~cm$^{-2}$ and (b) $\rho = 5 \times 10^{10}$~cm$^{-2}$, respectively. 
The results are obtained under constraints of a plunger gate correction below 20~mV, orbital energies above 1~meV in both dots, and tunnel gaps ranging between $10$ and $250$~$\mu$eV. 
The histograms in the top and right panels show the distributions of tunnel coupling and detuning correction, respectively. The dark (light) gray shading indicates the number of samples that fall inside (outside) the success box. The overlay step histogram showing the entire distribution, including all the data points.
 }
    \label{fig:yield}
\end{figure}

While the fluctuations of individual parameters are important, the ultimate test of scalability is how many devices in an array are functional. A single critical parameter falling out of its operational range can lead to unusable qubit. To capture this, we define the yield as the percentage of DQD instances that meet a set of realistic operational criteria after pre-tuning.

For a DQD to be considered functional, we demand that: (i) the orbital energy in each dot remains above 1 meV to suppress thermal excitations; (ii) the tunnel coupling lies in the range of $10-250$ $\mu$eV, which is required for high-fidelity gate operations; (iii) the barrier height above 0.5 meV to preserve DQD potential, and (iv) the initial detuning can be corrected with less than 20 mV of plunger voltage, limiting the required tuning range.

Our analysis reveals that the yield is critically affected by the charge density $\rho$. As shown in Fig.~\ref{fig:yield}, the yield decreases sharply with increasing disorder. For a more quantitative comparison, Figs.~\ref{fig:yield}~(a) and (b) show realizations of parameter samples for $\rho = 5 \times 10^{9}$~cm$^{-2}$ and $\rho = 5 \times 10^{10}$~cm$^{-2}$. The side histograms display the distributions of tunnel coupling and detuning correction, while the color scale represents the barrier height $h_B$. We observe that at lower density, the estimated yield is approximately 76\%, whereas at higher density, $\rho = 5 \times 10^{10}$~cm$^{-2}$, it decreases to 20\%. 

\begin{figure}[tb!]
    \centering
    \includegraphics[width=\columnwidth]{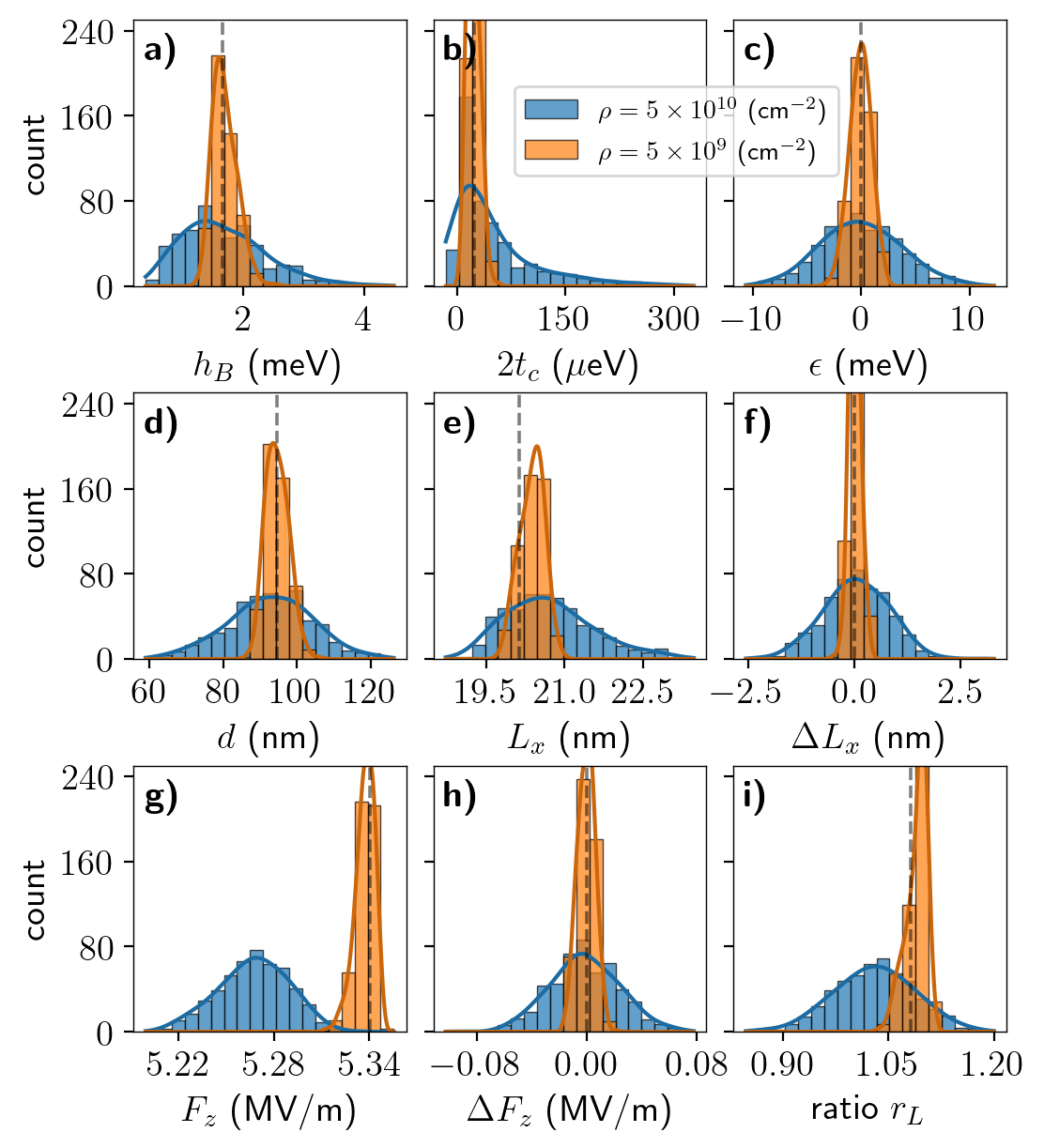}
    \caption{Marginal distributions of the key DQD parameters as function of charge density $\rho$.  The corresponding ensemble-averaged value of each DQD parameter are summarized  in Table~\ref{tab:avg_parameters}. The vertical dashed line represents the defect-free DQD parameters, serving as a reference for a device where the gates are set to \( V_{L/R} = 0.35 \, \text{V} \) and \( V_B = -0.1 \, \text{V} \). }
    \label{fig:hists}
\end{figure}

\begin{table}[tb!]
\centering
\caption{Statistical summary of parameters, including mean, standard deviation (std), and the coefficient of variation (CV) for two densities $\rho_1 = 5\times 10^{9}$~cm$^{-2}$ and $\rho_2 = 5\times 10^{10}$ cm$^{-2}$.}
\begin{tabular*}{\columnwidth}{@{\extracolsep{\fill}} l c c c c @{}}
\hline\hline
Parameter & & Mean ($\mu$) & Std ($\sigma$) & CV ($\sigma/|\mu|$)\\
\hline
\multirow{2}{*}{$d$ (nm)} & $\rho_1$ & 94.60 & 3.10 & 0.033\\
    & $\rho_2$ & 92.75 & 11.23 & 0.12\\
\multirow{2}{*}{$h_B$ (meV)} & $\rho_1$ & 1.691 & 0.203 & 0.12\\
    & $\rho_2$ & 1.678 & 0.692 & 0.41\\
\multirow{2}{*}{$\epsilon$ (meV)} & $\rho_1$ & $-0.072$ & 0.987 & --\\
    & $\rho_2$ & 0.036 & 3.683 & --\\
\multirow{2}{*}{$F_z$ (MV/m)} & $\rho_1$ & 5.337 & 0.006 & 0.001\\
    & $\rho_2$ & 5.267 & 0.022 & 0.004\\
\multirow{2}{*}{$\Delta F_z$ (MV/m)} & $\rho_1$ & $<10^{-4}$ & 0.006 & --\\
    & $\rho_2$ & $<10^{-3}$ & 0.024 & --\\
\multirow{2}{*}{$\Delta L_x$ (nm)} & $\rho_1$ & 0.009 & 0.139 & --\\
    & $\rho_2$ & 0.049 & 0.761 & --\\
\multirow{2}{*}{$L_x$ (nm)} & $\rho_1$ & 20.37 & 0.235 & 0.012\\
    & $\rho_2$ & 20.709 & 0.837 & 0.040\\
\multirow{2}{*}{$2t_c$ ($\mu$eV)} & $\rho_1$ & 23.13 & 9.13 & 0.40\\
    & $\rho_2$ & 50.30 & 57.91 & 1.15\\
\multirow{2}{*}{$r_L$, [$r_R$]} & $\rho_1$ & 1.090, [1.090] & 0.0146, [0.0144] & 0.013, [0.013]\\
    & $\rho_2$ & 1.030, [1.030] & 0.0561, [0.0566] & 0.054, [0.055]\\
\hline\hline
\end{tabular*}
\vspace{0.5ex}
\parbox{\columnwidth}{\footnotesize\raggedright
\textit{Note.}  CV is not informative when the mean is close to zero or changes sign (e.g., $\epsilon$, $\Delta F_z$, $\Delta L_x$); the large values reflect the instability of $\sigma/|\mu|$. For these quantities, variability is better characterized by $\sigma$. Means slightly offset from zero likely arise from finite-size and numerical-accuracy effects.}
\label{tab:avg_parameters}
\end{table}

\subsection{Fluctuations in Key Operational Parameters}
To quantify the contribution of each individual parameter to the overall variability, we now examine the variability of each DQD parameter separately. Figure~\ref{fig:hists} shows the marginal distributions for key parameters at two charge densities. While many distributions are approximately Gaussian, parameters like the tunnel coupling ($t_c$), barrier height ($h_B$) and electric field ($F_z$) exhibit significant non-Gaussian tails, including a finite probability of extreme variations that can compromise device function.

As shown in Figure~\ref{fig:hists}, an increase in charge density generally widens the parameter distributions. Higher densities increase the probability of smaller interdot distances $d$, creating a much heavier tail in the distribution for tunnel couplings $t_c$. As detailed in Table~\ref{tab:avg_parameters}, this leads to more than a threefold increase of the standard deviation of $d$, resulting in a six-fold increase of standard deviation of $t_c$.

Crucially, while shifts in mean values can often be compensated for by tuning gate voltages, we find that the relative variability, $\sigma/\mu$, consistently increases with charge density for all parameters. This highlights a fundamental challenge, as the intrinsic randomness becomes more pronounced at higher disorder levels.

\subsection{Tunnel Coupling Fluctuations}
\begin{figure}[htb!]
    \centering
    \includegraphics[width=0.8\columnwidth]{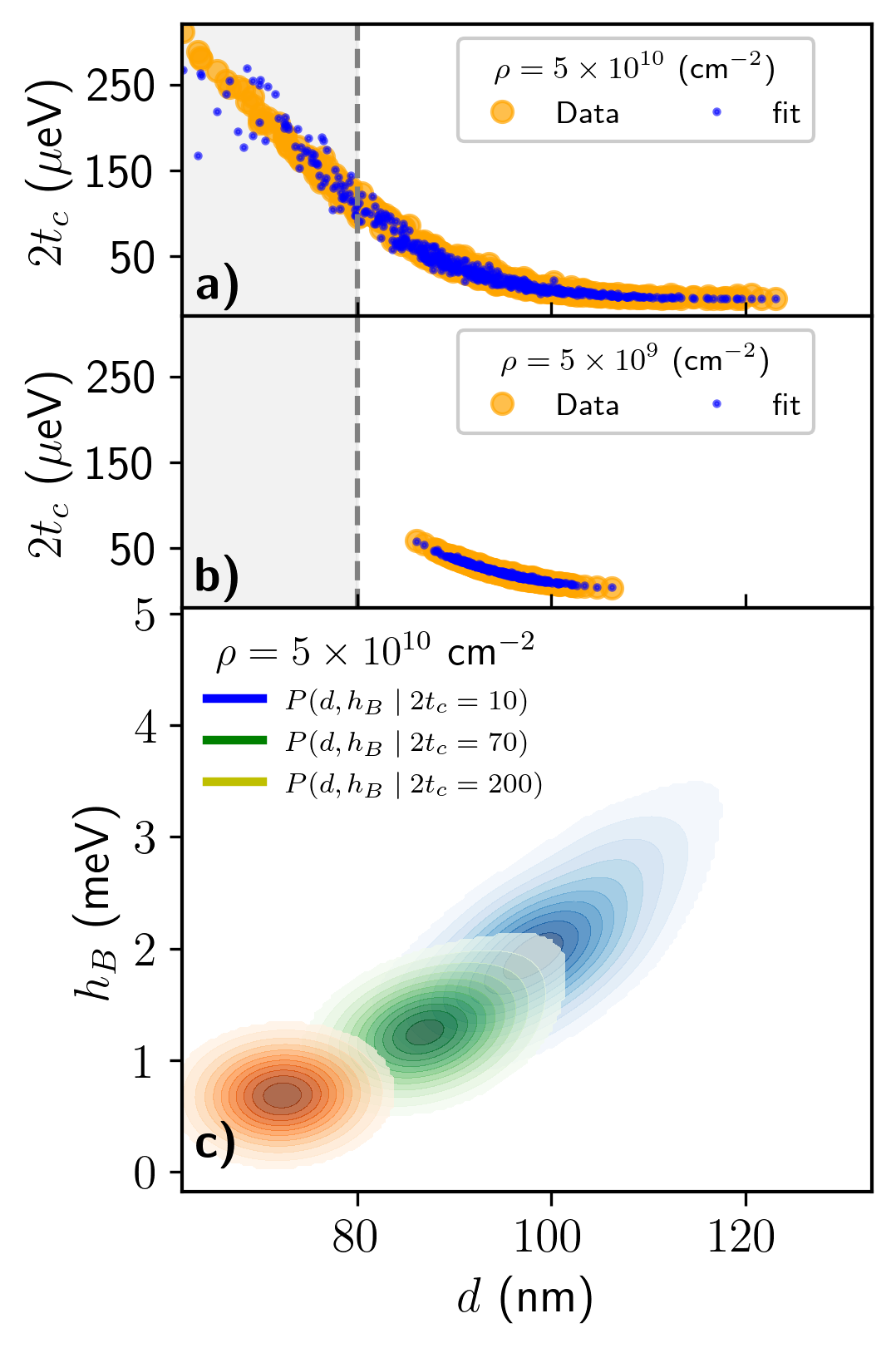}
    \caption{Fluctuations in the tunnel coupling $t_c$ as a function of interdot distance $d$ for two charge densities: (a) $\rho = 5 \times 10^{10}\,\mathrm{cm}^{-2}$ and (b) $\rho = 5 \times 10^{9}\,\mathrm{cm}^{-2}$. (c) Joint conditional density $P(d,h_B \mid 2t_c =\mathrm{value})$ for three values of $t_c$. The results indicate that variations in $t_c$ are primarily driven by fluctuations in $d$, which is positively correlated with $h_B$.}
    \label{fig:wkb}
\end{figure}
We now focus on the non-Gaussian parameter, tunnel coupling $t_c$, which is critical for the speed and fidelity of two-qubit gates \cite{Xue_Nature22}. According to the WKB approximation \cite{Wu2022processor_wkb,Cheng2025noise_wkb}, $t_c$ depends exponentially on both the inter-dot distance $d$ and barrier height $h_B$: $t_c = t_{c,0}e^{-\beta d\sqrt{2 m h_B} /\hbar}$. Thus even small fluctuations in $d$ and $h_B$ can lead to large variations in $t_c$, as shown by the first-order approximation:
\begin{equation}
    \delta t_c \approx -t_c\frac{\beta\sqrt{2m}}{\hbar} \left(\sqrt{h_B} \, \delta d +  \frac{d}{2\sqrt{h_B}}\delta h_B\right).
\end{equation}
Hence the variability in tunnel coupling, quantified as $\langle \delta t_c^2 \rangle$, depends on the variances of both $d$ and $h_B$, as well as their covariance, $\langle \delta d \, \delta h_B \rangle$.

Figure~\ref{fig:wkb}~(a)-(b) shows the simulated tunnel coupling $t_c$ as a function of inter-dot distance $d$ for two charge disorder densities. For both densities, we observe a strong, non-linear correlation consistent with the WKB approximation, confirming that as disorder pushes the dots closer together, the tunnel coupling increases dramatically. The higher disorder density in panel (a) can push some devices into a small-separation ($d < 80$ nm) regime where the WKB approximation breaks down. However, the strong correlation between $t_c$ and $d$ is preserved. This suggests that the fluctuations in tunnel coupling can be primarily explained by the fluctuations in $d$, which is positively correlated with $h_B$ as shown in Fig.~\ref{fig:wkb}c. Thus influence of $h_B$ can be effectively absorbed into a renormalized coefficient $\beta'$ in $\delta t_c \sim -t_c\frac{\beta'}{\hbar} \sqrt{h_B}  \delta d$, and allows the barrier height variations to be excluded from our subsequent analysis of disorder modes.

The strong correlation between $t_c$ and $d$ indicates that measuring $t_c$ fluctuations across an ensemble of devices can serve as a direct probe of variations in their physical separation. Furthermore, disorder-induced increases in $d$ can lead to a severe suppression of the tunnel coupling, significantly impairing spin qubit operations. This issue is particularly detrimental for applications like bucket-brigade charge shuttling \cite{Langrock_PRXQ23,Krzywda_PRB25}, in which an electron is supposed to be adiabatically transferred by global voltage pulses along a chain of tunnel-coupled quantum dots. The probability of unsuccessful charge shuttling, and the amount of spin dephasing of the shuttled qubit, is expected to be dominated by the interdot transfer across weak links, i.e.~pairs of adjacent QDs having particularly low values of $t_c$ \cite{Krzywda_PRB20,Krzywda_PRB21,Krzywda_PRB25}.
For example, for $\rho = 5 \times 10^{9}$ cm$^{-2}$ ($5\times 10^{10}$ cm$^{-2}$), we can observe a collapse of the tunnel gap of $2t_c \le 10\,\mu$eV in 6\% (25\%) of our simulated cases. The difference can be attributed to the large variance at larger densities, which on one hand increases the mean, but at the cost of an increasing probability of sampling an outlier. Dealing with such a weakly tunnel-coupled pair of quantum dots requires individual tuning that will make long distance shuttling resource intensive. In contrast, the relatively small variance gives an additional possibility of global retuning of the quantum dot towards typically shallower, more coupled dots, without inducing unwanted outliers. As demonstrated in Appendix~\ref{app:pca}, at lower defect charge density $\rho = 5 \times 10^{9}$~cm$^{-2}$ one can adjust the tuning point dimish the fraction of DQDs with $2t_c < 10\,\mu$eV to 2.5\%, as an effect of moving the average to larger values, while keeping the variance in a reasonable range, which is impossible at larger densities.

\section{\label{sec:PCA} Uncovering the Structure of Disorder}

Having established the impact of disorder on key operational parameters, we now move beyond analyzing marginal distributions and uncover the underlying structure of the device-to-device variability. The correlations between parameters visualised in the corner plot in Fig.~\ref{fig:generative}, contain crucial information about the common physical origin of the fluctuations. In this section, we first construct a predictive model of the disordered device. We then use principal component analysis (PCA) to deconstruct the DQD parameter randomness due to disorder noise into its most physically meaningful components.

\subsection{The predictive model}

To capture the full statistical behavior of the DQD, we model the joint distribution of the parameter vector $\mathbf{X} = [d, 2t_c, L_x, \Delta L_z, F_z, \Delta F_z, \epsilon]^T$ as a multivariate normal distribution. The model is fully specified by the mean vector $\boldsymbol{\mu}$ and the covariance matrix $\boldsymbol{\Sigma}$, which are estimated directly from our ensemble of simulated devices:
\begin{gather}\label{eq:norm_dist}
    P(\mathbf{X}) = \frac{1}{\sqrt{(2\pi)^k |\boldsymbol{\Sigma}|}} \exp\left(-\frac{1}{2}(\mathbf{X}-\boldsymbol{\mu})^T \boldsymbol{\Sigma}^{-1} (\mathbf{X}-\boldsymbol{\mu})\right), \\
    \boldsymbol{\mu} = \langle \mathbf{X} \rangle, \quad 
    \boldsymbol{\Sigma} = \langle (\mathbf{X} - \boldsymbol{\mu})(\mathbf{X} - \boldsymbol{\mu})^T \rangle.
\end{gather}
This parameterized distribution, $P(\mathbf{X}|\boldsymbol{\mu}, \boldsymbol{\Sigma})$, serves as a predictive statistical model, capable of generating realistic artificial device data by sampling from multivariate normal distribution.

\begin{figure}
    \centering
    \includegraphics[width=1.0\columnwidth]{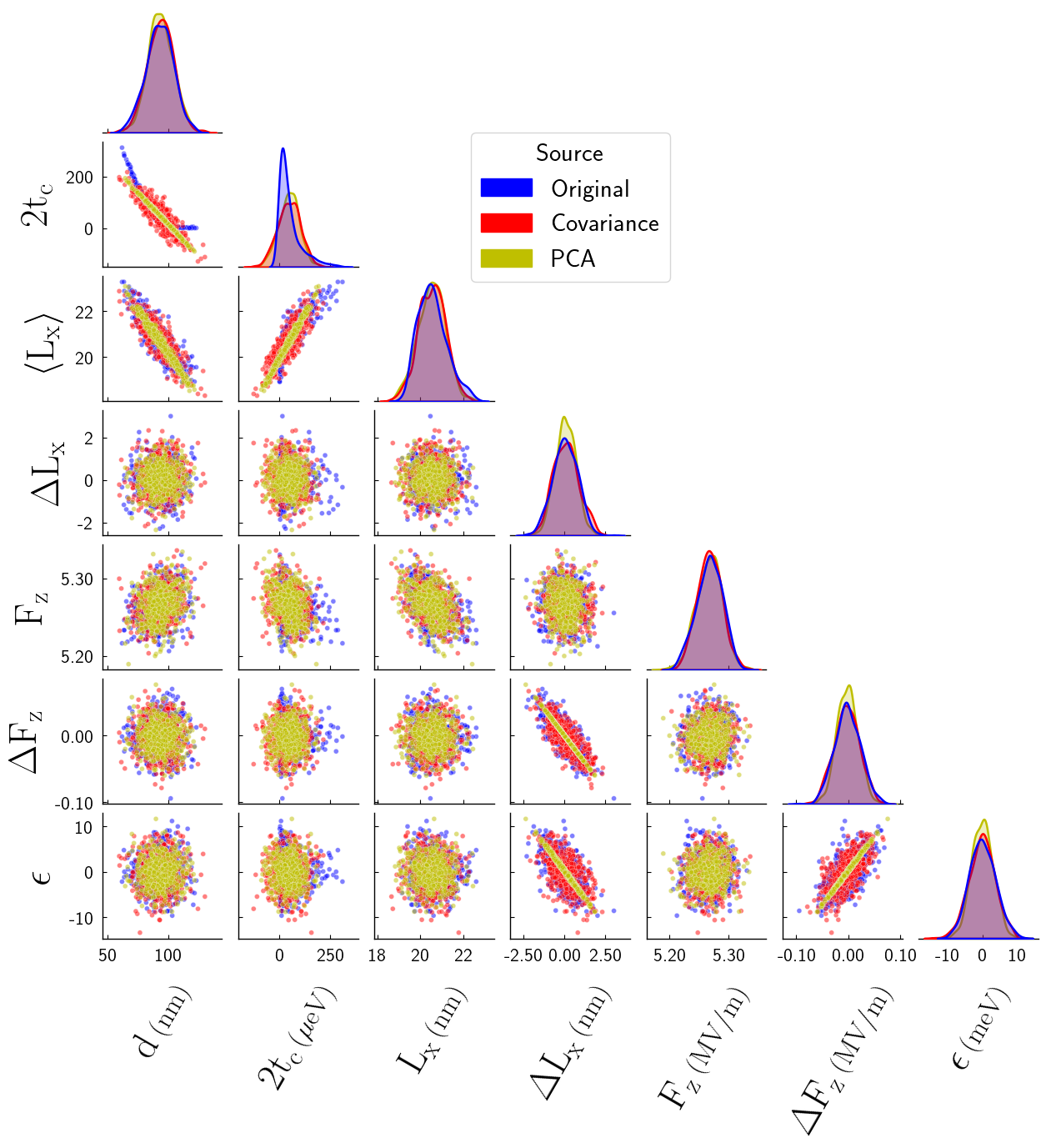}
    \caption{\textbf{Model validation via synthetic sampling}. A scatter plot matrix showing pairwise correlations between key DQD parameters. Comparison of (i) a multivariate normal model (red) and (ii) a reduced principal-component model using the first three components (yellow) against the reference dataset from full simulations (blue). The diagonal cells show overlays of the three 1D marginals, demonstrating close agreement with the original data. All three data (original and reconstructed) are obtained for density $\rho = 5 \times 10^{10}$ cm$^{-2}$.}
    \label{fig:generative}
\end{figure}

To validate this model, we generate an artificial dataset by sampling from the model and compare it to the original data from our full physical simulations. In Fig.~\ref{fig:generative}, a corner plot overlaying the two datasets demonstrates agreement in both the 1D histograms and the 2D correlations. Our statistical model is a high-fidelity representation of the complete device ensemble, with small discrepancies observed in the marginal distribution of $F_z$ and previously analyzed exponential relation between $2t_c$ and $d$. 

\subsection{Principal Component Analysis}

While the covariance matrix $\boldsymbol{\Sigma}$ contains all information about the correlations, its structure is not immediately interpretable. To identify the dominant, independent axes of variation, we perform principal component analysis (PCA) on the dimensionless correlation matrix,
\begin{equation}
    \mathbf{C}_{ij} = \frac{\boldsymbol{\Sigma}_{ij}}{\sqrt{\boldsymbol{\Sigma}_{ii} \boldsymbol{\Sigma}_{jj}}}.
\end{equation}
By solving the eigenvalue problem $\mathbf{C} \mathbf{d}_i = \lambda_i \mathbf{d}_i$, we obtain the eigenvectors $\mathbf{d}_i$ (the principal components) and their corresponding eigenvalues $\lambda_i$, which can be interpreted as the variance captured by each component \cite{Carleo_RMP19}.

The eigenvalue spectrum is shown in Fig.~\ref{fig:pca}~(a). The variance is heavily concentrated in the first few components, with the first three PCs accounting for over 80\% of the total variability. The remaining components fall within the random region predicted by the Marchenko-Pastur law \cite{marvcenko1967distribution}. As a result we assume any realization of disorder can be effectively approximated as a linear combination of just these three principal components.

To prove the point, in Fig.~\ref{fig:generative} we have also plotted the artificial data generated by sampling the noise only along the first three principal components. The close agreement with the full dataset confirms that the essential correlated structure shared by both the real and artificial data corresponds to this three-dimensional subspace defined by the dominant principal components.

\subsection{Physical Interpretation of Disorder Modes}
\begin{figure}[htb!]
    \centering
    \includegraphics[width=0.99\columnwidth]{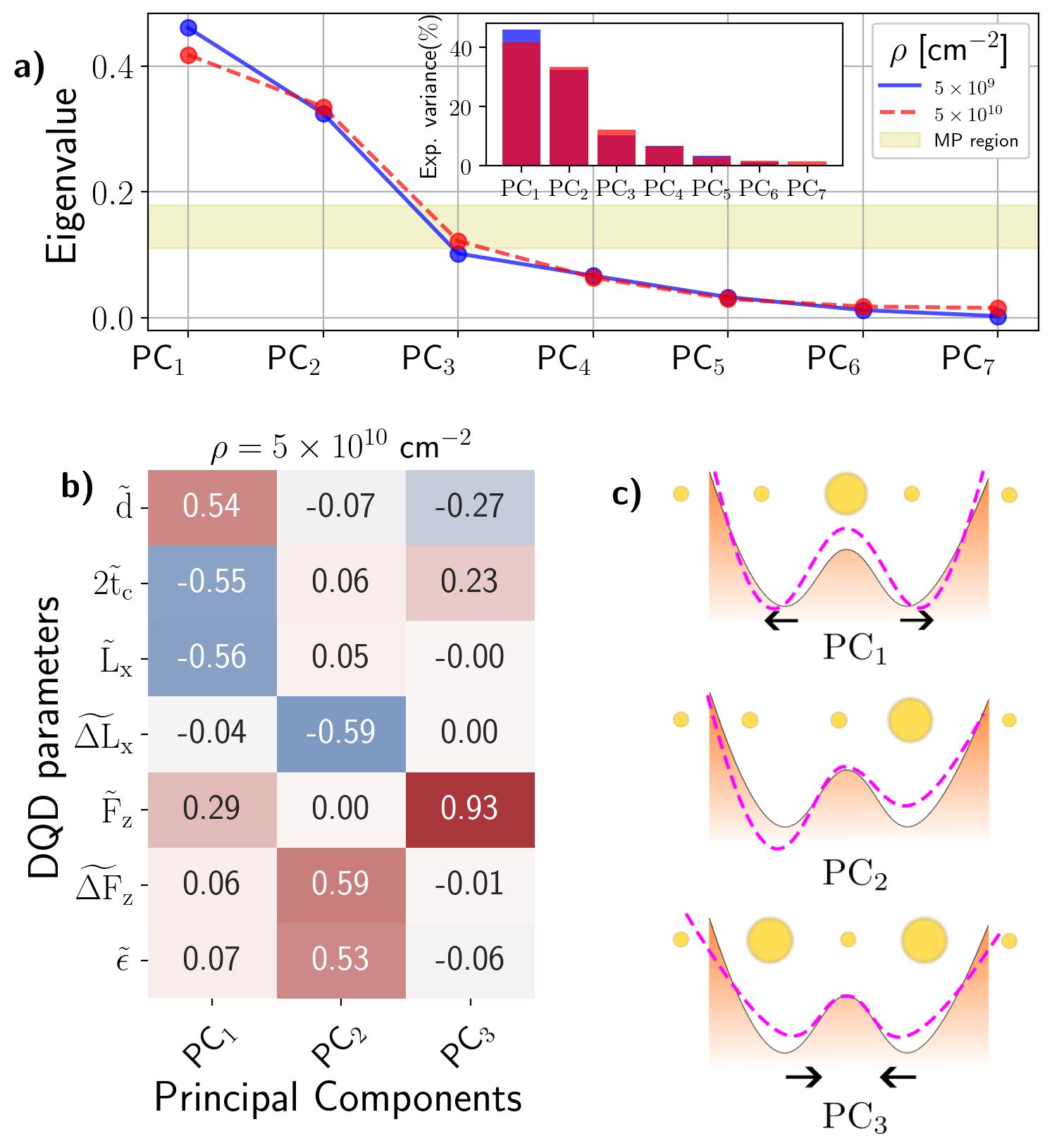}
    \caption{\textbf{Parameter Correlations from Disorder.} 
    (a) The eigenvalues corresponding to each PC for two densities, $5\times 10^{9}$ cm$^{-2}$ (blue) and $5\times 10^{10}$ cm$^{-2}$  (red). Inset:  Corresponding explained variance of each PC.  (b) Eigenvectors of the first three PCs, consistent with the schematic in (c).  The datasets are chosen for fixed plunger gate $V_{L/R}=0.35$ V and barrier gate $V_B=-0.1$ V. }
    \label{fig:pca}
\end{figure}

We can understand the physical nature of these dominant modes by inspecting their composition, i.e. the elements of the eigenvectors, which are shown in Fig.~\ref{fig:pca}~(b). Each component represents a specific, correlated way in which the DQD potential is deformed by the charge disorder, as illustrated schematically in Fig.~\ref{fig:pca}~(c).

To be precise, in the first mode the disorder-induced increase of $d$ is correlated with decrease of $2t_c$, decrease of confinement length $x$ direction for both dots, and increase in the $F_z$ electric field for both dots. 
This represents a symmetric squeezing or stretching of the DQD potential. An excess of negative charge located between the dots pushes the dots apart, leading to a decrease of $2t_c$.

The second mode is dominated by the fluctuations of energy detuning $\epsilon$ and the differential electric field $\Delta F_z$. This corresponds to an asymmetric tilting of the double-well potential, which directly affects the detuning. This mode is primarily caused by excess charges located closer to one dot than the other.

The third significant mode is almost entirely composed of the average vertical field $F_z$. This represents a common-mode shift in the potential of both dots relative to the central barrier. It is distinct from PC$_1$,  as it has a weaker effect on the lateral confinement, and primarily reflects a change in the average vertical electric field, which is important for valley splitting \cite{Friesen_PRB07,Saraiva_PRB09}.

\begin{figure*}
    \centering
    \includegraphics[width=1\linewidth]{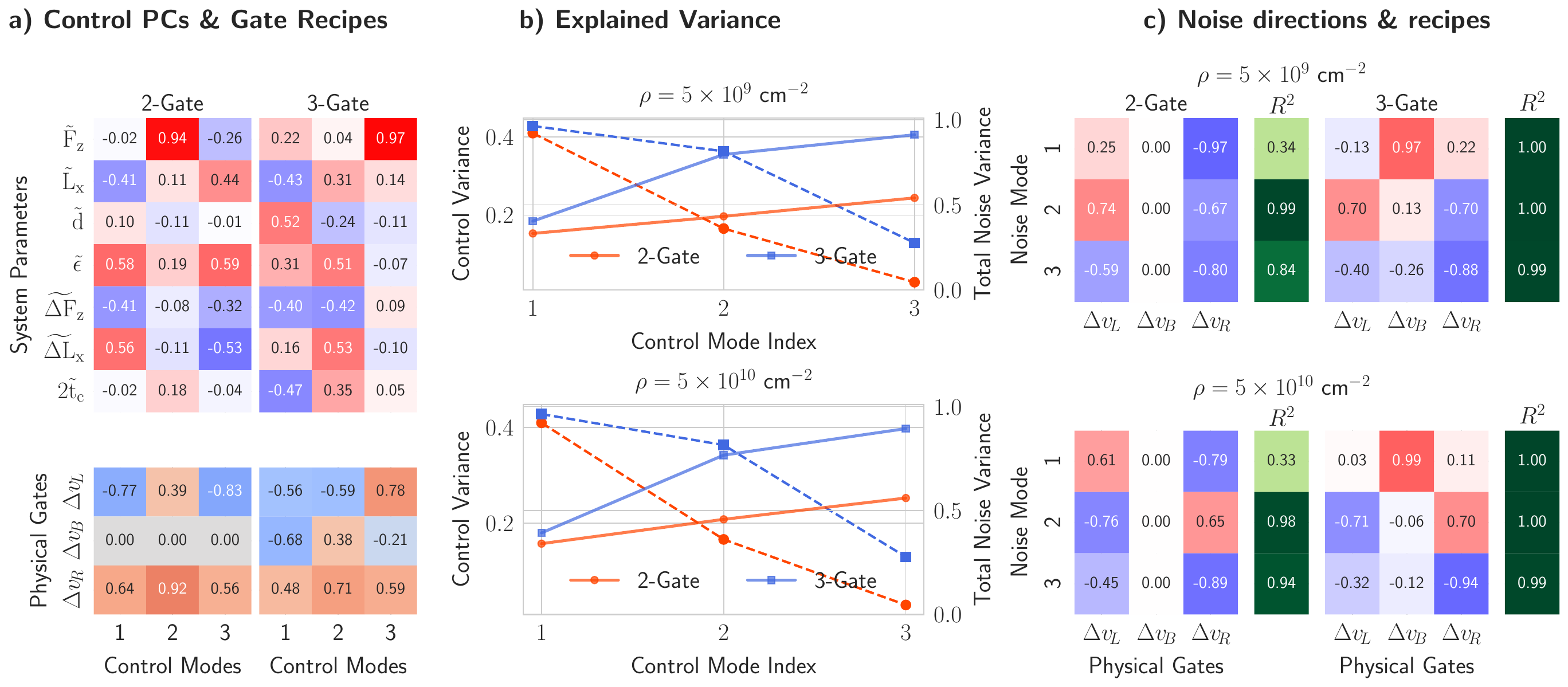}
    \caption{\textbf{Comparison of two- and three-gate control schemes}. (a) Composition of control modes in gate-voltage space and determination of control directions, $\boldsymbol{\beta}_j^c$, via least-squares regression.  (b) Controllability, $\eta_K$, representing the fraction of disorder variance explained by the  $K$ control modes. The three-gate control (blue line) accounts for over 90\% of the variance, whereas the plunger-only control (red line) captures about 50\%. (c) Derived voltage recipes, $\boldsymbol{\beta}^d_j$, and corresponding $R^2_j$ values for two densities: (i) $5\times 10^{9}$ cm$^{-2}$ and (ii) $5\times 10^{10}$ cm$^{-2}$. The disorder data for both densities, used in this figure, were obtained after applying the tuning procedure. Additionally, a comparison between the pre-tuning and post-tuning results is shown in Fig.~\ref{fig:control_1e10}.}
    \label{fig:control}
\end{figure*}

\begin{figure}[htb!]
    \centering
    \includegraphics[width=0.9\columnwidth]{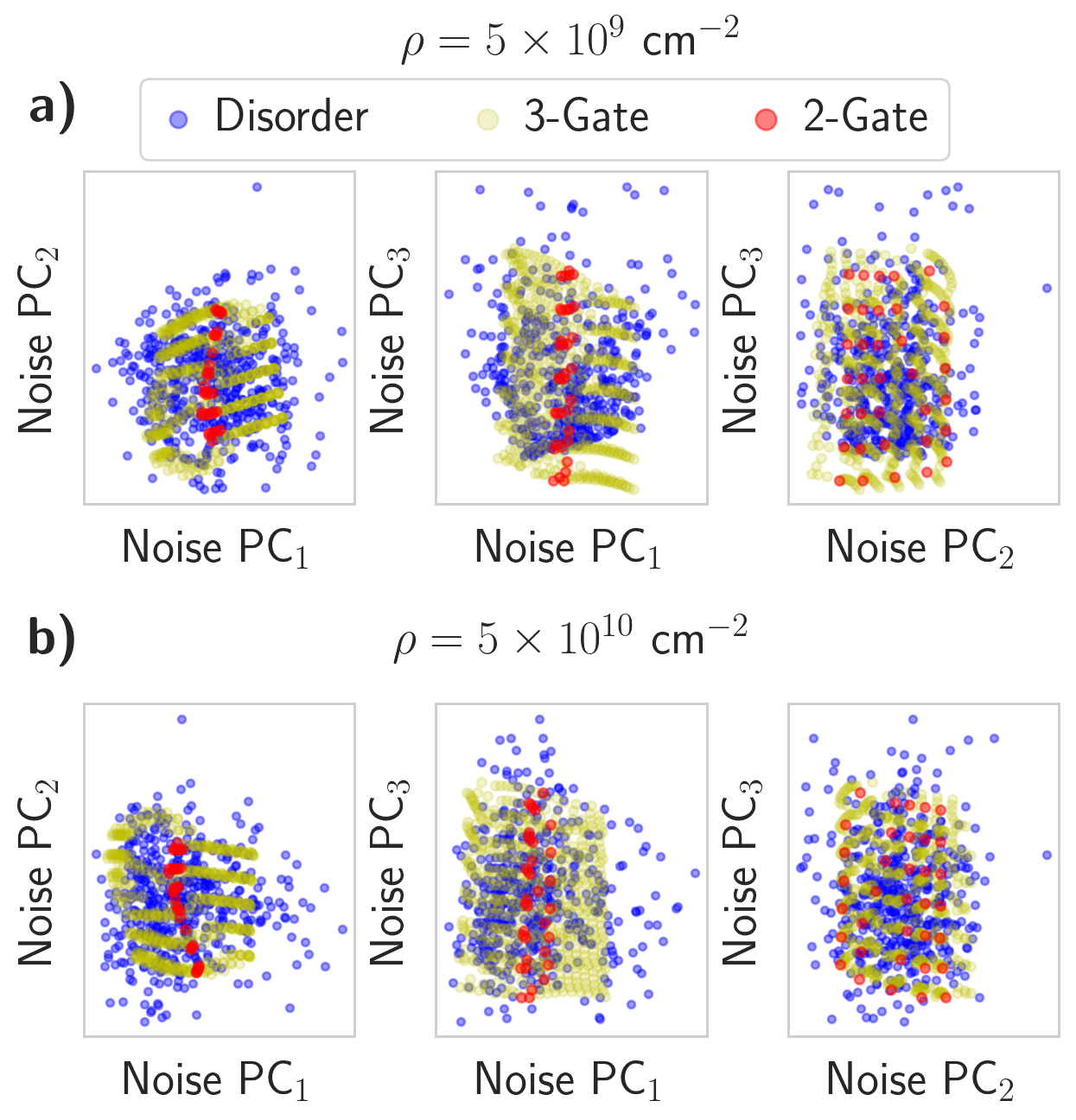}
    \caption{Projected two-gate (red) and three-gate (yellow) control data compared with disorder data (blue) for two densities $5\times 10^{9}$ and $5\times 10^{10}$, showing that three-gate control spans the full disorder space, while two-gate control is confined to an approximately two-dimensional subspace.}
    \label{fig:projPCs}
\end{figure}

\section{Application: Disorder-Aware Controllability \label{sec:control}}
We now apply principal component analysis (PCA) to quantify the system controllability. We define controllability as the ability to mitigate the effects of disorder by tuning the gate voltages. Specifically, we investigate whether a more scalable plunger-only control scheme \cite{Ivlev2025} can mitigate disorder as effectively as the recently more common approach that uses both plunger and barrier gates.

\subsection{Control Modes}
First, we identify the control modes: the primary directions in the system's parameter space that can be accessed by tuning gate voltages. To do this, we generate a control dataset in the absence of disorder. Starting from a set of initial parameters, we sweep the gate voltages across their operational ranges: the barrier gate $V_B \in (-90, -110)$~mV,  the average plunger voltage ${V_P = (V_L + V_R)/2 \in (330, 380)}$~mV, and the plunger voltage difference $\Delta V = V_L - V_R \in (0, 20)$~mV. The $\Delta V$ sweep is kept asymmetric ($\Delta V \geq 0$) to avoid introducing additional nonlinearities. This process yields a matrix of system parameters, $\mathbf{X}_{3g} = \mathbf{X}(V_P, \Delta V, V_B)$, where each row corresponds to a specific gate voltage configuration.
We compare here the full 3-gate control with a plunger-only (2-gate) control scheme. For the latter, the barrier gate is held fixed, yielding the parameter matrix $\mathbf{X}_{2g} = \mathbf{X}(V_P, \Delta V, V_B = \text{const})$.

We standardize both datasets to create a matrix $\mathbf{\tilde{X}}$, where each column has a mean of zero and a standard deviation of one. PCA is then performed on $\mathbf{\tilde{X}}$ to find the principal control modes, an orthonormal basis of vectors which we denote as $\mathbf{c}_j$. The top panel of Figure~\ref{fig:control}(a) shows the composition of these modes in the parameter space. Next, we determine the direction in the gate voltage space, $\boldsymbol{\beta}_j^c$, corresponding to each principal mode using linear regression. Specifically, each $\boldsymbol{\beta}_j^c$ is found by solving the following least-squares problem:
\begin{equation}
    \boldsymbol{\beta}_j^c = \underset{\boldsymbol{\beta}}{\operatorname{argmin}} \sum_{i} \left[ (\tilde{\mathbf{x}}_i \cdot \mathbf{c}_j) - (\Delta \mathbf{V}_i \cdot \boldsymbol{\beta}) \right]^2
\end{equation}
Here, $\Delta \mathbf{V}_i = \mathbf{V}_i - \langle \mathbf{V} \rangle$ is the gate voltage vector for data point $i$ centered around the dataset's mean voltage $\langle \mathbf{V} \rangle$, and $\tilde{\mathbf{x}}_i$ is the standardized system parameter vector. The resulting unit vectors in voltage space, $\boldsymbol{\beta}_j^c / \lVert\boldsymbol{\beta}_j^c\rVert$, define the principal control directions and are given in the bottom panel of Fig.~\ref{fig:control}(a). 

As shown in Fig.~\ref{fig:control}(a) and (b), the number of statistically relevant control modes match the number of independent gate voltages, resulting in two for plunger-only control and three for 3-gate control. This one-to-one correspondence indicates that the system's response to gate voltages is approximately linear. It also suggests that the 2-gate control spans a two-dimensional manifold in the parameter space, which may be insufficient to compensate for higher-dimensional effects of disorder.

By definition, the voltage-space vectors $\boldsymbol{\beta}_j$ represent the optimal strategies for inducing maximum variation along their corresponding principal modes. For most control modes, this involves the simultaneous change of all available parameters. A notable exception is the third mode in the 3-gate scheme, which primarily creates strong, symmetric variations in the z-component of the electric field at the dot locations, leaving other system parameters largely unaffected.

\subsection{Controllability}
We now quantitatively check to what extent these control capabilities can counteract the device variablity caused by the disorder.
The effects of disorder are captured by a set of PCA vectors, the disorder modes $\{\mathbf{d}_j\}$, demonstrated in Fig.~\ref{fig:pca}. We restrict our analysis to the first three disorder modes, which capture over 90\% of the total variance.

To illustrate how well the control space covers the disorder space, we project the control data points onto the subspace spanned by the first three disorder modes calculated for two densities. The results, shown in Fig.~\ref{fig:projPCs}, compare the projected 2-gate (red) and 3-gate (yellow) control data against the projected disorder data (blue). It is evident that the 3-gate control spans all dimensions of the disorder space, while the 2-gate control is confined to an approximately two-dimensional subspace. In particular, the plunger-only control struggles to compensate for the first disorder mode, as the plane spanned by the red points is nearly orthogonal to the first disorder axis (noise PC$_1$).

Figure~\ref{fig:control}~(b) quantifies this by plotting the controllability (right axis), $\eta_K$, defined as the fraction of disorder variance explained by the first $K$ control modes:
\begin{equation}
    \eta_K = \frac{\sum_{j=1}^K \text{Var}(\mathbf{\tilde{Y}} \cdot \mathbf{c}_j)}{N} \,\, .
    \label{eq:explained_variance}
\end{equation}
Here, $\mathbf{\tilde{Y}}$ is the standardized disorder dataset and $N$ is the number of parameters. While the 3-gate control (blue line) can account for over 90\% of the disorder variance, the plunger-only control (red line) corresponds to only about 50\%, with most of that captured by its first control mode alone.

\subsection{Optimal Compensation Recipes}
Finally, we derive explicit gate voltage recipes to compensate for each disorder mode. We begin by decomposing each disorder mode $\mathbf{d}_j$ onto the basis of control modes $\mathbf{c}_i$:
\begin{equation}
    \mathbf{d}_j \approx \sum_i \alpha_{ij} \mathbf{c}_i
\end{equation}
The coefficients $\alpha_{ij} = (\mathbf{d}_j \cdot \mathbf{c}_i)/(\mathbf{c}_i \cdot \mathbf{c}_i)$ quantify how much each control mode $\mathbf{c}_i$ contributes to the disorder mode $\mathbf{d}_j$. The quality of this reconstruction is measured by the coefficient of determination, $R^2_j$, defined as
\begin{equation}
R^2_j = 1 - \frac{\| \mathbf{d}_j - \sum_i \alpha_{ij} \mathbf{c}_i \|^2}{\| \mathbf{d}_j \|^2} \,\, ,
\end{equation}
where a value close to 1 indicates that the disorder's effects can be controlled. 

An explicit gate recipe to counteract the j-th disorder mode is then the corresponding linear combination of the control mode recipes, $\boldsymbol{\beta}_i^c$:
\begin{equation}
    \boldsymbol{\beta}^d_j = \sum_i \alpha_{ij} \boldsymbol{\beta}_i^c
\end{equation}

Figure~\ref{fig:control}(c) shows the derived voltage recipes, $\boldsymbol{\beta}^d_j/\abs{\boldsymbol{\beta}^d_j} = (\Delta v_{Lj}^d, \Delta v_{Bj}^d, \Delta v_{Rj}^d)$, and their corresponding effectiveness, as quantified by the $R_j^2$ coefficient. The analysis quantitatively confirms the limitations of plunger-only control, reflected in the low $R^2$ value of 0.30 for the first disorder mode, which accounts for nearly half of the total variance. In contrast, the 3-gate control can effectively compensate for all three disorder modes, with $R^2$ values approaching 1.

Crucially, the derived voltage recipes are physically intuitive. While both control schemes can compensate for the second disorder mode (primarily a detuning shift) and the third mode (a common-mode shift), their responses to the first disorder mode are very different. For 3-gate control, this strong disorder in the inter-dot potential can be effectively compensated by changing the barrier gate voltage—an action unavailable in the plunger-only scheme.

Our framework demonstrates robustness not only against the overall disorder strength (various densities $\rho$) but also against shifts in the pre-set gate voltages ($V_{L/R}$ and $V_B$) that define the initial operating point and the parameters of the dot in the disorder-free structure. While the shapes of marginal distributions of the key DQD parameter do depend on the choice of these voltages, the PCA results, i.e.~the structure of correlations between the DQD parameters, and the conclusions concerning the controllability of DQDs, are robust to the choice of them. This is illustrated in Appendix~\ref{app:pca}, confirming the wide applicability of our findings to Si/SiGe DQDs. 

We have also repeated in Appendix~\ref{app:notuning} the above data analysis procedure to compare the controllability before and after the pre-tuning procedure---the detuning correction required to obtain a symmetric DQD. This dataset was post-selected based on yield conditions to avoid pathological cases, such as a single elongated dot. Crucially, the controllability results remain consistent with the finely-tuned case, confirming the model's predictive power.

\section{Discussion and conclusion} \label{sec:discussion}
Our analysis demonstrates that charge disorder in Si/SiGe double quantum dots, while a source of significant device-to-device variability, does not lead to mutually independent fluctuations of DQD parameters. Instead, it induces highly structured variations of these parameters along specific modes. These modes can be identified with certain real-space arrangements of charges in the oxide that lead to correlated fluctuations in the DQD parameters.

Beyond this specific insight, our work establishes the Principal Component Analysis (PCA) as a powerful framework for analysis of variability of properties of semiconductor nanostructures that host spin qubits.
Based on spectral analysis of the normalized covariance matrix, it provides a systematic tool for analyzing multivariate data. In particular, PCA allowed us to decompose the disorder fluctuations into three dominant modes: a symmetric squeezing/stretching of the DQD potential, an asymmetric tilting of the double-well potential, and a common-mode shift in the potential of both dots relative to the central barrier. These three modes account for more than 90\% of the total variance in the DQD parameters and allow for a clear physical interpretation of the effect of the disorder.

By quantifying these correlations, we have constructed a multivariate Gaussian model, which is fully specified by the mean and covariance matrix. In this way, we have developed a predictive model that can be used to generate realistic artificial data without the need for computationally expensive electrostatic simulations, thereby facilitating the development of advanced tuning algorithms.

By characterizing the amplitude of the fluctuations, we have also analyzed how the variability changes as a function of charged defect density $\rho$ over an order-of-magnitude wide range of $\rho$. We have confirmed that reproducibility, as quantified by the device yield drops as disorder amplitude increases. We have analysed the marginal distributions of the key DQD parameters, and found that while many of them are approximately Gaussian, the tunnel coupling specifically shows non-Gaussian features. We have related the variability of the tunnel coupling to fluctuations in the inter-dot distance, which we have found to be strongly correlated with the barrier height. This observation suggests that a measurement of $t_c$ can serve as a direct probe of the inter-dot distance, with possible verification through measuring spin splittings of two single-spin qubits in the two dots in presence of a magnetic field gradient \cite{Watson_Nature18,Zajac_Science18,Xue_Nature22}. We have also analyzed the implications of this tunnel coupling variability for charge shuttling, finding that for a charge density above $\rho = 5 \times 10^{9}$ cm$^{-2}$ ($\rho = 5\times 10^{10}$ cm$^{-2}$), at least 6\% (25\%) of the dot pairs in a chain would require resource-intensive, individual barrier gate tuning. 

As a novel in context of control of quantum dots application of PCA, we have used it to quantify the limitations of plunger-only control. We have compared the three-mode PCA decomposition of the disorder space with the PCA decomposition of the control space, which was obtained by sweeping the gate voltages in the absence of disorder. A comparison of the manifolds spanned by the most important PCs allowed us to visualize the overlap between the control and disorder spaces, and to quantify system controllability, defined as the ability to mitigate the effects of disorder by tuning gate voltages. We have found that while 3-gate control can explain over 90\% of the disorder variance, plunger-only control explains only about 50\%.

For practical purpose of mitigating these disorder modes, we have derived explicit gate voltage recipes to compensate for each one. The results confirm that plunger-only control is severely limited in its ability to counteract the first, dominant disorder mode, which accounts for nearly half of the total variance of DQD parameters. This stands in stark contrast to three-gate control, which can effectively compensate for all three modes with recipes that directly match their physical interpretation. While plunger-only control can address the second mode (a detuning shift) and the third mode (a common-mode vertical shift), it struggles with the first mode. The symmetry of the two plunger gates makes it difficult to compensate for changes in $d$ (and hence in $t_c$) without simultaneously altering the potential symmetry and introducing unwanted changes in $\epsilon$, $\Delta L_x$, and $\Delta F_z$.

This suggests that while plunger-only control is a more scalable approach, it may require more sophisticated strategies. For instance, it might be necessary to introduce additional control knobs by relying on cross-talk from neighboring plunger gates. Alternatively, one could operate in a non-linear regime where the manifold spanned by the two control modes is a curved surface that could potentially cover more of the disorder space. Of course, one can try to correct for a specific parameter, for instance $t_c$, by tuning the plunger gates in a symmetric way. However, our analysis suggests that such an approach will lead to a simultaneous change in the electric field $F_z$ while failing to correct the change in $d$. The former might adversely modify the valley physics in Si/SiGe qubits and the electric control of spin qubits in all platforms, while the latter leaves a primary issue for two-qubit gates and the operation of singlet-triplet qubits unresolved.

It should be noted that since we focus on electrostatic disorder only, we neglect the randomness of valley coupling values (and thus of valley splitting values) in the two dots resulting from the roughness of the Si/SiGe interface and the interdiffusion of Ge atoms \cite{Wuetz_NC23,Losert_PRB23,Lima_MQT23,Klos_AS24,Thayil_arXiv24}. Unequal values of complex valley coupling parameter in the two dots are known to affect the tunnel coupling between the valley and orbital ground states in the two dots \cite{Zhao_PRAPL22,Tariq_NPJQI22,Cywinski2025_inprep}. Our neglect of this effect means that the results presented in this paper for the randomness of $t_c$ due to electrostatic disorder apply to the case of atomic disorder at Si/SiGe interface resulting in the ``deterministic'' regime of valley couplings \cite{Losert_PRB23,Volmer_NPJQI24}. In the case of currently often encountered Si/SiGe dots with valley coupling in the ``disordered'' regime \cite{Losert_PRB23,Volmer_arXiv25}, the tunnel coupling $t_c$ calculated here is related to the inter- and intra-valley tunnel couplings $t_{\pm}$ by $t_c^2 \! = \! |t_+|^2+|t_{-}|^2$ \cite{Zhao_PRAPL22,Tariq_NPJQI22,Cywinski2025_inprep}.
The atomic disorder at Si/SiGe interface  also affects any properties of Si/SiGe spin qubits that are connected with the effective spin-orbit coupling in a given dot, such as the electron $g$-factors \cite{Ruskov_PRB18,Woods_arXiv24} and spin-valley couplings \cite{Yang_NC13,Volmer_NPJQI24,Volmer_arXiv25}. While the knowledge of variability of parameters such as confinement length, dot ellipticity, and electric fields in the $z$ direction that affect the overlap of the electron wavefunction with the interface, will be relevant for full characterization of variability of $g$-factors and spin-valley couplings, including the atomistic disorder at the interface will be crucial for these properties. 

Concluding, we have analyzed the device-to-device variability in Si/SiGe double quantum dots arising from electrostatic disorder and developed a predictive statistical model to generate reliable artificial data for tuning algorithm development. Using the Principal Component Analysis on a large simulated dataset, we have found that the resulting parameter fluctuations were taking place along a few directions in parameter space. We have identified three such fluctuation modes, with the most significant one being caused by fluctuations in number of defect charges localized in-between the dots. This mode leads to correlated fluctuations in the inter-dot distance and barrier height that can fully explain the strong variation in the tunnel coupling. By quantifying the system controllability, we have shown that such a mode cannot be fully corrected by a plunger-only control scheme, which lacks the necessary barrier control. Our work establishes PCA as a powerful framework for understanding and mitigating disorder in quantum dot devices, and highlights the importance of barrier control for scalable qubit operation.

\section*{Code and Data Availability}
The data and code used to produce the results in this study are available online \cite{qdot-disorder-structure}.

	\begin{acknowledgments}
	    This work was funded by the European Union’s Horizon Research and Innovation Actions under Grant Agreement No. 101174557 (QLSI2). J.A.K acknowledges funding from Dutch National Growth Fund (NGF) as part of the Quantum Delta NL programme. The authors thank Evert van Nieuwenburg for insightful conversations on PCA.

	\end{acknowledgments}

\appendix

\section{PCA and fluctuations for different pre-set gate voltages ($V_{L/R}=0.19$ V and $V_B=-0.06$ V)}\label{app:pca}

\begin{figure}[htb!]
    \centering
    \includegraphics[width=\columnwidth]{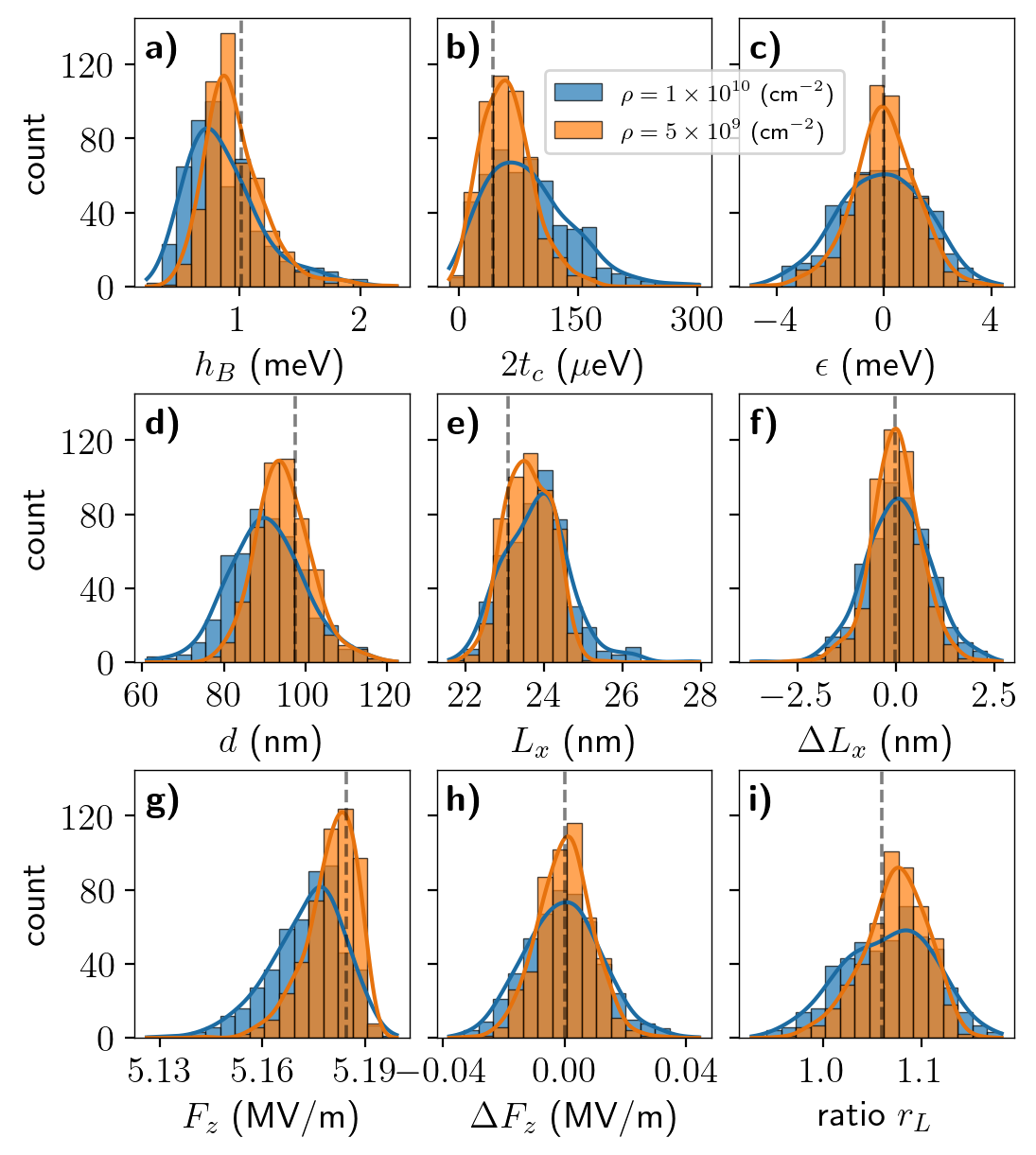}
    \caption{Marginal distributions of the key DQD parameters as function of charge density $\rho$.  The corresponding ensemble-averaged value of each DQD parameter are summarized  in Table~\ref{tab:avg_parameters_app}.
    The vertical dashed line indicates the nominal defect-free DQD baseline (set by $V_0=0.19$ V and $V_B=-0.06$ V). The key parameters for this baseline are: interdot distance $d_0 = 97$ nm, barrier height $h_B = 1.01$ meV, and tunneling gap $2 t_{c,0} \approx 42\,\mu\text{eV}$. The dots are elliptical with lateral dimensions $L_{x,0} \approx 24$ nm and $L_{y,0} \approx 22$ nm (ratio $r_D=1.09$) with average electric field $F_z\approx 5.18$ MV/m. }
    \label{fig:hists_app}
\end{figure}

\begin{table}[h!]
\centering
\caption{Statistical summary of parameters, including mean, standard deviation (std), and the coefficient of variation (CV) for two densities $\rho_1 = 5\times10^{9}$ and $\rho_2 = 1\times10^{10}$ cm$^{-2}$. The datasets are chosen for fixed plunger gate $V_{L/R}=0.19$ V and barrier gate $V_B=-0.06$ V. }
\begin{tabular*}{\columnwidth}{@{\extracolsep{\fill}} l c c c c @{}}
\hline\hline
Parameter & & Mean ($\mu$)& Std ($\sigma$) & CV ($\sigma/|\mu|$)\\
\hline
\multirow{2}{*}{$d$ (nm)} & $\rho_1$ & 94.67 & 6.67 & 0.07\\
    & $\rho_2$ & 90.54 & 9.01 & 0.09\\
\multirow{2}{*}{$h_B$ (meV)} & $\rho_1$ & 0.96 & 0.25 & 0.26\\
    & $\rho_2$ & 0.87 & 0.31 & 0.35\\
\multirow{2}{*}{$\epsilon$ (meV)} & $\rho_1$ & 0.02 & 1.17 & --\\
    & $\rho_2$ & $-0.09$ & 1.58 & --\\
\multirow{2}{*}{$F_z$ (MV/m)} & $\rho_1$ & 5.18 & $0.007$ & 0.001\\
    & $\rho_2$ & 5.17 & $0.01$ & 0.002\\
\multirow{2}{*}{$\Delta F_z$ (MV/m)} & $\rho_1$ & $< 10^{-4}$ & $0.008$ & --\\
    & $\rho_2$ & $<10^{-4}$ & $0.012$ & --\\
\multirow{2}{*}{$\Delta L_x$ (nm)} & $\rho_1$ & $<10^{-3}$ & 0.63 & --\\
    & $\rho_2$ & 0.05 & 0.81 & --\\
\multirow{2}{*}{$L_x$ (nm)} & $\rho_1$ & 23.60 & 0.58 & 0.02\\
    & $\rho_2$ & 23.81 & 0.82 & 0.03\\
\multirow{2}{*}{$2t_c$ ($\mu$eV)} & $\rho_1$ & 60.96 & 31.70 & 0.52\\
    & $\rho_2$ & 87.01 & 53.33 & 0.60\\
\multirow{2}{*}{$r_L$, [$r_R$] } & $\rho_1$ & 1.07, [1.07] & 0.03, [0.03] & 0.03, [0.03]\\
    & $\rho_2$ & 1.06, [1.06] & 0.045, [0.046] & 0.042, [0.043]\\    
\hline\hline
\end{tabular*}
\vspace{0.5ex}
\parbox{\columnwidth}{\footnotesize\raggedright
\textit{Note.}  CV is not informative when the mean is close to zero or changes sign (e.g., $\epsilon$, $\Delta F_z$, $\Delta L_x$); the large values reflect the instability of $\sigma/|\mu|$. For these quantities, variability is better characterized by $\sigma$. Non-zero means ($\mu$) likely arise from finite-size and numerical-accuracy effects. 
}
\label{tab:avg_parameters_app}
\end{table}

To further test the robustness of our conclusions and the applicability of the PCA approach across different regimes, we performed the same analysis using for another set pre-tuned values for the plunger voltages ($V_{L/R} = 0.19\text{ V}$) and the gate barrier voltage ($V_{B} = -0.06\text{ V}$).
We perform this analysis to verify that the results for the PCA and controllability are not sensitive to the choice of voltages defining the shape and properties of DQD in a disorder-free structure, whereas the fluctuations of key DQD parameters depend on the gate voltages and charge densities. Let us note that with the voltages chosen here the double-well character of the electrostatic potential in absence of disorder is less pronounced than for parameters used in the main text, and as a result the DQD would vanish for many realizations of charge disorder for the highest $\rho \! =\! 5 \times 10^{10}$ cm$^{-2}$ considered in te paper. Consequently, we focus here on lower densities, $\rho_1 = 5 \times 10^{9} $ cm$^{-2}$ and $\rho_2 = 1 \times10^{10} $ cm$^{-2}$.

In Fig.~\ref{fig:hists_app}  the corresponding marginal distributions of key parameters for two charge densities are shown. Similarly, while many of the parameters follow approximately Gaussian distributions, certain quantities—such as tunnel gap (\(2t_c\)), barrier height (\(h_B\)), and vertical electric field (\(F_z\)) again show significant non-Gaussian behavior. 

For these two considered densities that  differ by a factor of two, the broadening of the higher-density distribution relative to the lower one is less pronounced compared to the cases shown in the main text (see Fig.~\ref{fig:hists}), where the densities there differ by an order of magnitude. Nevertheless, the difference remains observable. The details, including the mean, standard deviation, and coefficient of variation, are presented in Table~\ref{tab:avg_parameters_app}.

\begin{figure}[htb!]
    \centering
    \includegraphics[width=0.99\columnwidth]{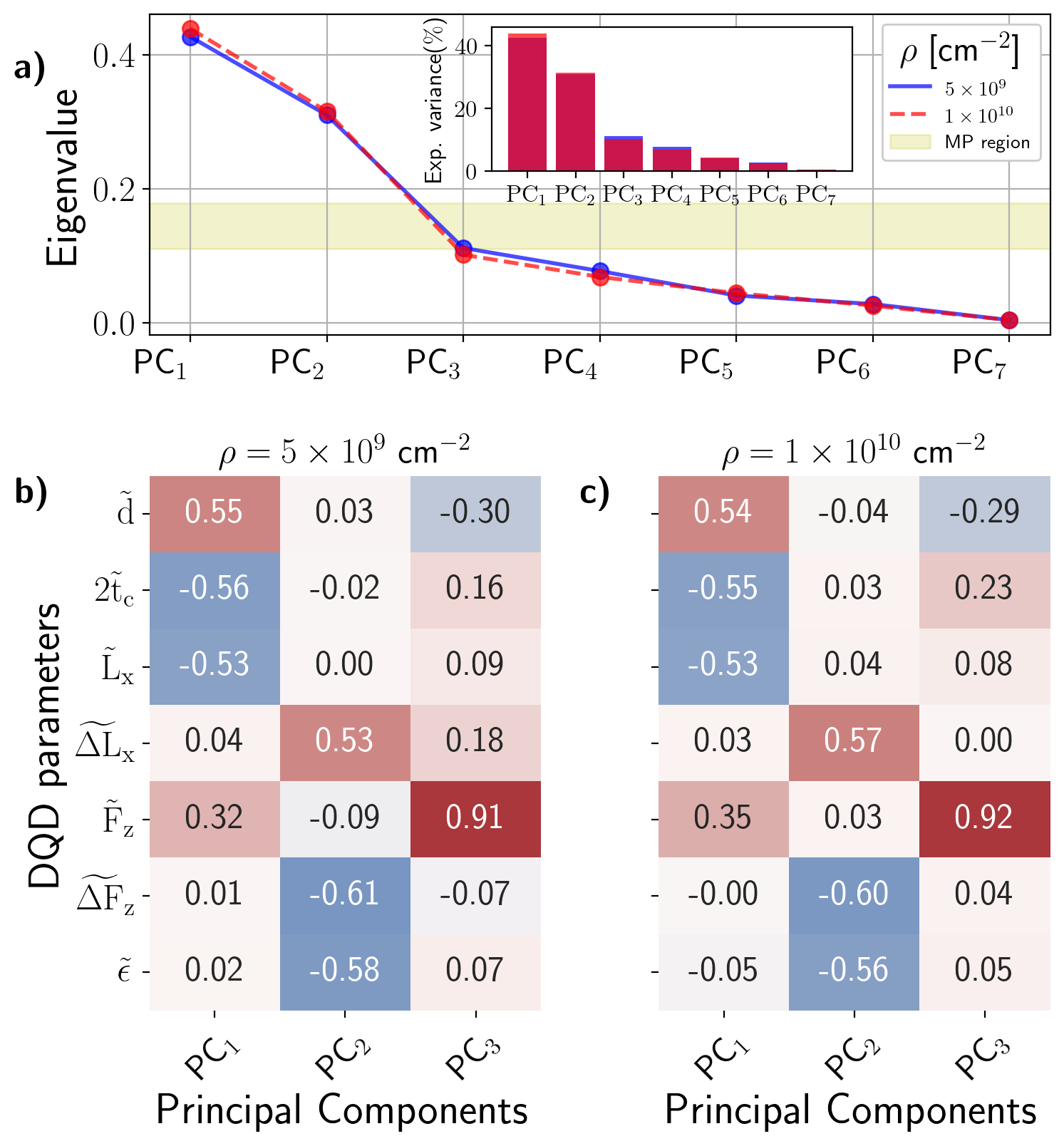}
    \caption{\textbf{Parameter Correlations from Disorder.} 
    (a) The eigenvalues corresponding to each PC for two densities, $5\times 10^{9}$ cm$^{-2}$ (blue) and $1\times 10^{10}$ cm$^{-2}$  (red). Inset:  Corresponding explained variance of each PC.  (b)-(c) Eigenvectors of the first three PCs.  The datasets are chosen for fixed plunger gate $V_{L/R}=0.19$ V and barrier gate $V_B=-0.06$ V.  }
    \label{fig:pca_app}
\end{figure}

Crucially, the results of the PCA calculations for this new voltage setting are quantitatively very similar to those for the previous one, suggesting that the nature of these dominant modes is primarily related to the spatial distribution of defects rather than the initially pre-set gate voltage values. The resulting eigenvalue spectra and eigenvector compositions, presented in Fig.~\ref{fig:pca_app}, show that the variance remains predominantly concentrated in the first few principal components, with the first three accounting for over 80\% of the total variability—closely matching the behavior observed at the reference density in Fig.~\ref{fig:pca}. Moreover, the spatial profiles of the dominant eigenmodes exhibit similar patterns of correlated potential deformation, reinforcing that the fundamental nature of disorder-induced fluctuations is largely insensitive to charge density within this range.

\section{Controllability results for DQD without pre-tuning} \label{app:notuning}

Using the modified gate settings defined in Appendix~\ref{app:pca}, we apply the same PCA-based approach to quantify the controllability for a dataset at a density of $\rho=1 \times 10^{10}\,\text{cm}^{-2}$. In this way we benchmark our approach in two ways. First, we have changed the tuning point to analyze the controllability of shallower dots at smaller charge densities. Second, we analyze two distinct scenarios: one dataset with the typical pre-tuning (detuning correction) applied, and one without any detuning correction. This benchmarking is crucial because the influence of charged defects strongly depends on their density $\rho$ and the pre-set gate voltages, especially when working with shallower (less confined) dots. 

In contrast to the main text, here we operate at shallower-dots conditions ($V_{L/R}=0.19\,\mathrm{V}$, $V_{B}=-0.06\,\mathrm{V}$), which allow to observe DQDs for densities of $\rho \! =\! 5 \times 10^{9}\,\mathrm{cm}^{-2}$ and $\rho \!=\! 1 \times 10^{10}\,\mathrm{cm}^{-2}$, but not at the higher $\rho \! =\! 5 \times 10^{10}\,\mathrm{cm}^{-2}$ considered in the main text. At these lower densities, the defects mainly introduce a modest shift and deformation in the electrostatic potential, which still allows the two distinct quantum dot minima to remain clearly visible for most disorder realizations. We highlight that at higher densities ($\rho=5 \times 10^{10}\,\text{cm}^{-2}$), the defect influence is strong enough to completely wash out one or both dots, often resulting in shallow or single-dot potentials, which are far from the desired operating point. Overcoming this issue requires significant adjustment of the gate voltages, often demanding stronger initial confinement - and this is what we have done in the main text. For shallower dots, we again rely on sweeping the gate voltages across their operational ranges: the barrier gate $V_B \in (-60, -50)$~mV, the average plunger voltage $V_P = (V_L + V_R)/2 \in (180, 210)$~mV, and the plunger voltage difference $\Delta V = V_L - V_R \in (0, 10)$~mV.

\begin{figure*}
    \centering
    \includegraphics[width=1\linewidth]{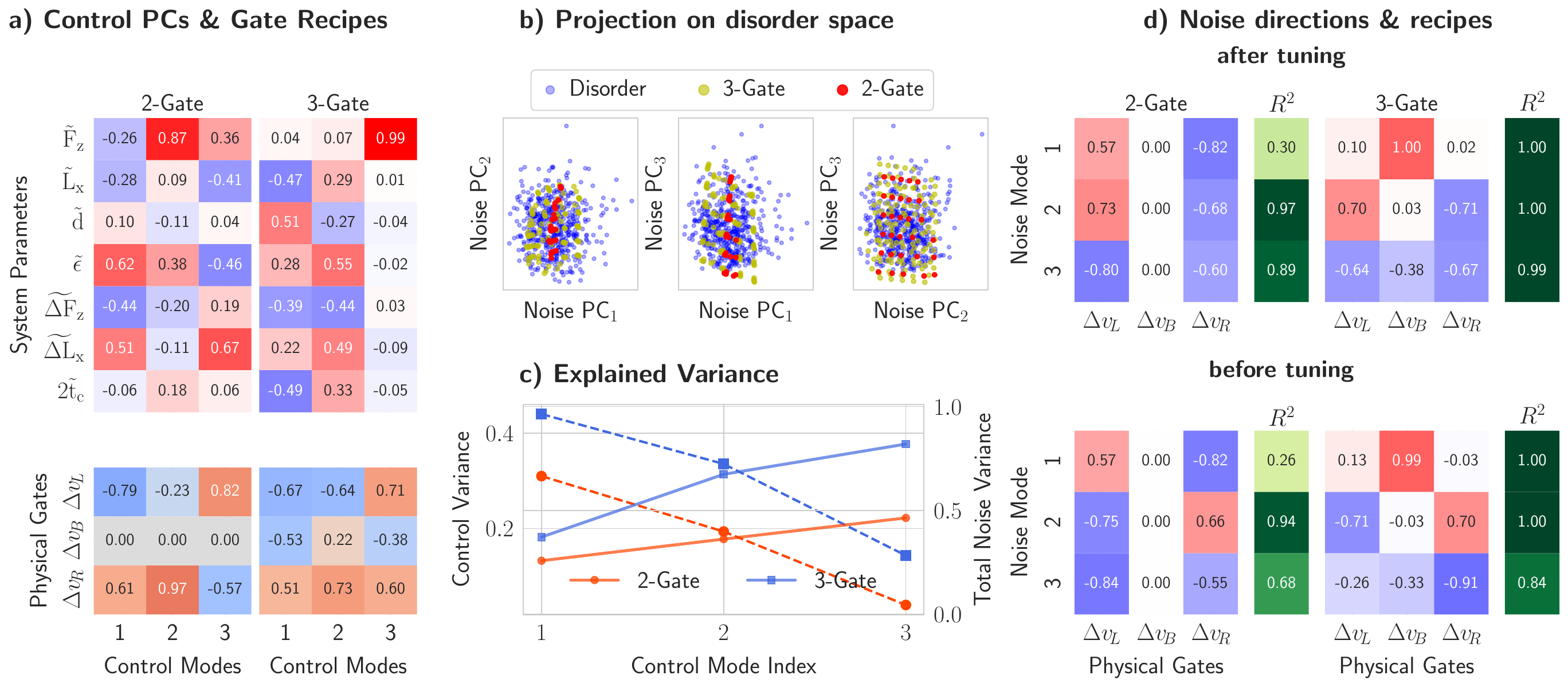}
    \caption{\textbf{Comparison of two- and three-gate control schemes}. (a) Composition of control modes in gate-voltage space and determination of control directions, $\boldsymbol{\beta}_j^c$, via least-squares regression. (b) Projected two-gate (red) and three-gate (yellow) control data compared with disorder data (blue), showing that three-gate control spans the full disorder space, while two-gate control is confined to an approximately two-dimensional subspace. (c) Controllability, $\eta_K$, representing the fraction of disorder variance explained by the r $K$ control modes. The three-gate control (blue line) accounts for over 90\% of the variance, whereas the plunger-only control (red line) captures about 50\%. (d) Derived voltage recipes, $\boldsymbol{\beta}^d_j$, and corresponding $R^2_j$ values for (i) post-tuning and (ii) pre-tuning datasets. The disorder data is chosen for density $\rho = 1 \times 10^{10}$ cm$^{-2}$.  The datasets are chosen for fixed plunger gate $V_{L/R}=0.19$ V and barrier gate $V_B=-0.06$ V. }
    \label{fig:control_1e10}
\end{figure*}

The controllability results using dataset after detuning correction shown in Fig.~\ref{fig:control_1e10} are very similar to those discussed in Sec. \ref{sec:control}. This similarity is not surprising, given the almost identical control modes and PCA.
The consistency of these results is preserved when examining the dataset before detuning correction (see bottom panel of Fig.~\ref{fig:control_1e10}~(d)). Here, the voltage recipes and their corresponding effectiveness ($R^2$) for the first three control modes are similar to those obtained after tuning, except for the third mode which is slightly asymmetric, most likely due to more non-linear effects. Our analysis demonstrates that working in the regime in which both of the dots are well defined, while being detuned by a few $\text{meV}$, yields control analysis consistent with the one obtained at adjusted operating points.


\newpage 
\bibliography{ref,refs_Si,refs_charge_noise}

\end{document}